\documentclass{aastex631}
\usepackage{float}
\usepackage{rotating}

\shorttitle{Magnetized Disks with Outflows for CL-AGNs}
\shortauthors{Wu \& Gu}

\begin{document}

\title{Magnetized Accretion Disks with Outflows for Changing-look AGNs}

\author{Wen-Biao Wu}
\author{Wei-Min Gu}

\affiliation{Department of Astronomy, Xiamen University, Xiamen,
Fujian 361005, P. R. China; guwm@xmu.edu.cn}

\begin{abstract}
Changing-look active galactic nuclei (CL-AGNs) challenges the standard
accretion theory owing to its rapid variability. Recent numerical simulations 
have shown that, for the sub-Eddington accretion case, the disk is
magnetic pressure-dominated, thermally stable, and geometrically thicker than
the standard disk. In addition, outflows were found in the simulations.
Observationally, high blueshifted velocities absorption lines indicate that
outflows exist in AGNs. In this work, based on the simulation results, we
investigate the magnetic pressure-dominated disk, and find that the accretion
timescale is significantly shorter than that of the standard thin disk.
However, such a timescale is still longer than that of the CL-AGNs.
Moreover, if the role of outflows is taken into account,
then the accretion timescale can be even shortened.
By the detailed comparison of the theoretical accretion timescale with
the observations,
we propose that the magnetic pressure-dominated disk incorporating outflows
can be responsible for the rapid variability of CL-AGNs.
\end{abstract}


\section{Introduction} \label{sec: intro}
Active galactic nuclei (AGNs) have strong flux variations from infrared to X-rays 
and are typically classified into two types: type 1, which shows broad emission 
lines, and type 2, which does not. According to the AGNs unified model, these two 
types differ in their orientations with respect to the observer's line of sight 
to the dust torus surrounding the AGNs \citep[e.g.,][]{1993ARA&A..31..473A}. 
However, an increasing number of multi-wavelength observations (including X-ray \citep[e.g.,][]
{2017A&A...607L...9K,2018MNRAS.480.3898N,2019MNRAS.483L..88P,2021MNRAS.507..687J}, 
infrared \citep[e.g.,][]{2017ApJ...846L...7S,2018ApJ...862..109Y,2020MNRAS.491.4615K},
 and radio \citep[e.g.,][]{2016MNRAS.460..304K,2019NatAs...3..387P,
2021MNRAS.503.3886Y,2021MNRAS.506.4188L}) have identified a sub-class of AGNs 
referred to as changing-look AGNs (CL-AGNs), which are capable of transforming 
from type 1 to type 2 or vice versa over relatively short timescales. Due to 
the difficulties in explaining the timescale of CL-AGNs using the standard 
model \citep[][hereafter SSD]{1973A&A....24..337S}, it has become an increasingly 
interesting topic.

Several schemes have been proposed to explain the contradictions between theory 
and observation. The obscuration mechanism, based on the AGNs unified 
model \citep{1989ApJ...346L..21G} where obstructions enter or leave our line of 
sight, can cause the appearance or disappearance of broad emission lines. 
However, this scheme is ruled out due to significant 
variation of CL-AGNs' mid-infrared emissions and the absence of polarization or 
weak polarization \citep[e.g.,][]{2016MNRAS.457..389M,2017ApJ...846L...7S,2019A&A...625A..54H}. 
Another candidate is tidal disruption events \citep[e.g.,][]{2015MNRAS.452...69M,2016MNRAS.455.1691R,2020ApJ...898L...1R}, which predicts a change in the light curve with $t^{-5/3}$. 
Although it explains some cases, most observations do not support this model \citep{{2020ApJ...889...46S}}. 
It is generally accepted that these intrinsic changes from the accretion disk and 
variations in the accretion rate dominate this phenomenon. However, the viscosity 
timescale of the standard disk, which varies with the accretion rate, is much 
larger than that of the CL-AGNs \citep[e.g.,][]{2018ApJ...864...27S,2022arXiv221105132R}. 
Therefore, new models or modifications to the standard model have been proposed 
to address the issue. The role of the magnetic field in accretion disks has been 
studied and applied to the CL-AGNs \citep[e.g.,][]{2019MNRAS.483L..17D,2021ApJ...916...61F,2021MNRAS.502L..50S,2023arXiv230210226S}. The limit cycle of magnetically dominated disks or disk 
instabilities located in the narrow region between the inner ADAF and the outer 
standard disk is used to explain the repeating CL-AGNs
\citep[e.g.,][]{2020A&A...641A.167S,2021ApJ...910...97P,2022arXiv221100704K}. 
\cite{2020A&A...643L...9W} demonstrated that tidal interaction between disks in 
supermassive black hole binaries can also trigger CL-AGNs.

The standard thin disk model has served as the foundation
for understanding various astronomical 
phenomena since its proposal. However, it still faces challenges such as observations of 
cataclysmic variables and low-mass X-rays binaries, where the disk appears to be 
geometrically thicker than expected (see \cite{2007MNRAS.375.1070B} 
for details).
Most recently, numerical simulations \citep{2023ApJ...945...57H}
have shown that, for the sub-Eddington accretion case, the disk is dominated by 
the magnetic pressure and exhibits distinct properties from the standard thin
disk.
The magnetic pressure-dominated accretion disk is thermally stable,
geometrically thicker than the standard thin disk,
and has flatter effective temperature profiles. 
In addition, outflows were found in the simulations.
These features are consistent with \cite{2007MNRAS.375.1070B}'s
magnetically dominated viscous disks and some previous numerical simulations 
\citep[e.g.,][]{2016MNRAS.460.3488S,2016MNRAS.459.4397S,2019ApJ...885..144J} 
except for outflow. The two-dimensional radiation-magnetohydrodynamic numerical 
simulation by \cite{2011ApJ...736....2O} likewise discovers outflows 
driven by magnetic pressure on standard-type thin disks. 

Highly blueshifted absorption lines are frequently detected in the UV and X-ray 
spectra of AGNs \citep[e.g.,][]{2010A&A...521A..57T,2019A&A...627A.121S}, 
indicating the presence of high-velocity winds/outflows in these systems. These 
outflows, particularly ultrafast ones, are presumed to originate in the inner 
regions of accretion disks and could be powered by the magnetic field 
\citep[e.g.,][]{2021ApJ...914...31Y,2022MNRAS.513.5818W}. Recent observations 
have also validated the existence of outflows in CL-AGNs, such as MrK590 
\citep{2015ApJ...798....4G} and NGC1566 \citep{2021MNRAS.507..687J}. In this 
study, we utilized the results of numerical simulations to elucidate the CL-AGNs 
phenomenon by considering the impact of outflows. The remainder is organized as 
fellows. In Section $\ref{sec: model}$, we provide basic physics and equations 
for our model. Section $\ref{sec: numerical Results}$ shows numerical results and 
analyses. Section $\ref{sec: Summary}$ presents the conclusions and discussion. 

\section{Basic Equations} \label{sec: model}
For simplicity, a steady-state axisymmetric magnetized accretion disk with outflows
 is considered under the well-known pseudo-Newtonian potential 
$\Phi = -GM_{\rm{BH}}/(R-R_{\rm{g}})$ \citep{1980A&A....88...23P}, where 
$M_{\rm{BH}}$ is the mass of the black hole and $R_{\rm{g}} = 2GM_{\rm{BH}}/c^2$ 
is the Schwarzschild radius. Therefore, we assume the angular velocity is 
Keplerian, i.e., $\Omega = \Omega_{\rm{K}} = [GM_{\rm{BH}}/R(R-R_{\rm{g}})^2]^{1/2}$.

In this work, the magnetic pressure $P_{\rm{m}} = B^2/8\pi$, $B$ is the 
mean mid-plane magnetic field strength. The gas pressure 
$P_{\rm{g}} = \rho k_{\rm{B}} T_{\rm{c}}/\mu m_{\rm{p}}$, the radiation pressure 
$P_{\rm{r}} = aT_{\rm{c}}^4/3$, where $T_{\rm{c}}$ is the mid-plane temperature, 
$\rho$ is the mid-plane density, $\mu = 0.617$ is the mean molecular weight, 
$k_{\rm{B}}$ is the Boltzmann constant, and $a$ is the radiation constant. 
We adopt \cite{2007MNRAS.375.1070B} in assuming that the magneto-rotational 
instability-riven dynamo could amplify the turbulent (toroidal)
 field until saturation. 
The saturation is determined by the Alfv$\acute{\rm{e}}$n speed 
$V_{\rm{A}} \equiv \sqrt{P_{\rm{m}}/\rho}$ and roughly equals the geometric 
means of the Keplerian velocity $V_{\rm{K}}$ and the gas sound speed 
$c_{\rm{g}} = \sqrt{P_{\rm{g}}/\rho}$, i.e., 
\begin{equation}
V_{\rm{A}}^2 = \lambda V_{\rm{K}} c_{\rm{g}} \ ,\label{eq1}
\end{equation}
where we have introduced a parameter $\lambda$, which represents the 
strength of the turbulent (toroidal) magnetic field, and $\lambda$ is in the 
range $[0,1]$.
The condition $\lambda = 1$ corresponds to the saturation 
case of the turbulent (toroidal) magnetic field and has been observed by 
numerical simulations \citep[e.g.,][]{2023ApJ...945...57H}.

Due to the additional magnetic pressure, the total pressure $P =P_{\rm{m}}+ P_{\rm{g}}+P_{\rm{r}}$, 
the sound speed is written as 
\begin{equation}
c_{\rm{s}}^2 = \frac{P}{\rho} \ .\label{eq2}
\end{equation}

Numerical simulations have demonstrated the above magnetically dominated viscous disks
\citep[e.g.,][]{2016MNRAS.460.3488S,2016MNRAS.459.4397S,2019ApJ...885..144J,2023ApJ...945...57H}. 
The model shows a geometrically thicker disk and higher color temperatures, which 
would be stable against thermal and viscous instabilities. In addition, outflows 
were found in the simulations \citep{2023ApJ...945...57H}. The two-dimensional 
radiation-magnetohydrodynamic numerical simulation of \cite{2011ApJ...736....2O} 
also found outflows driven by magnetic pressure on standard-type thin disks 
(in their numerical simulation, the toroidal component of the magnetic field is 
dominant, i.e., $|B_{\varphi}|/B_{\rm{p}}>1$). Taking inspiration from the 
simulations above, we developed a new outflow model. The fundamental equations 
describing accretion disks with outflows employed in this study can be found in 
our previous work \citep{2022ApJ...930..108W}. The continuity equation is
\begin{equation}
\dot{M} =-2\pi R\Sigma V_{\rm{R}} \ , \label{eq3}
\end{equation}
where $V_{\rm{R}}$ is the inflow speed, which is defined to be negative when
the flow is inward, and $\Sigma$ is the surface density 
\begin{equation}
\Sigma = 2\rho H \ . \label{eq4}
\end{equation}

The scaleheight can be redefined based on the above analysis 
\begin{equation}
H = \frac{c_{\rm{s}}}{\Omega_{\rm{K}}}, \label{eq5}
\end{equation}
Similarly, we redefine the kinematic viscosity as
\begin{equation}
\nu = \alpha c_{\rm{s}}H, \label{eq6} 
\end{equation}
where $\alpha$ is the viscosity parameter.
In principle, $\alpha$ is related to the magnetization of the disk
\citep{2016MNRAS.460.3488S}. Thus, for high viscosity, such as $\alpha = 0.1$,
the parameter $\lambda$ may not be negligible.
The relationship between $\alpha$ and the magnetization of the disk is beyond
the scope of this paper. For simplicity, we take both $\alpha$ and $\lambda$
as free parameters in this work.

Due to the effect of outflow, we assume that the mass accretion rate varies with 
the radius \citep{1995A&A...295..807F,1999MNRAS.303L...1B} $\dot{M} = \dot{M}_{\rm{out}}
(R/R_{\rm{out}})^{p}$, where $\dot{M}_{\rm{out}}$ is the mass accretion rate at 
the outer boundary $R_{\rm{out}}$, in the present investigation, we set $p$ as 
a free parameter.

The azimuthal momentum equation is reduced to the algebraic form \citep{2022ApJ...930..108W}:
\begin{equation}
\nu \Sigma = \frac{\dot{M}fg^{-1}}{3\pi}\left(1-\frac{l^2p}{p+\frac{1}{2}}
\right) \ , \label{eq7}
\end{equation}
where $g = -\left(2/3\right)\left(d \ln \Omega_{\rm{K}}/d \ln R\right)$ and
the factor $f = 1-[\Omega_{\rm{K}} \left(3R_{\rm{g}}\right)/\Omega_{\rm{K}}(R)]\left(3R_{\rm{g}}
/R\right)^{p+2}$. $l$ is the parameter that characterizes the angular momentum 
taken away by the outflow \citep{1999MNRAS.309..409K}. In this study, 
we set $l > 1$ due to the presence of a strong magnetic field.

The energy equation is written as
\begin{equation}
Q_{\rm{vis}} = Q_{\rm rad} \ , \label{eq8}
\end{equation}
where $ Q_{\rm{vis}}$ and $ Q_{\rm rad}$ are the viscous
heating rate and the radiative cooling rate, respectively.
Their expressions are as follows \citep{2022ApJ...930..108W},

\begin{equation}
Q_{\rm{vis}} = \frac{3\dot{M}\Omega_{\rm{K}}^2fg}{4\pi}\left(1-\frac{l^2p}{p+\frac{1}{2}}\right) \\
              , \label{eq9}
\end{equation}

\begin{equation}
Q_{\rm rad} = \frac{16\sigma T_{\rm{c}}^4}{3\tau} \ , \label{eq10}
\end{equation}
where $\tau = \kappa_{\rm{es}} \rho H$ represents the optical depth, 
with $\kappa_{\rm{es}} = 0.34 \rm{\ cm^2 \ g^{-1}}$ being the electric 
scattering opacity. In the inner region of the accretion disk, electric 
scattering opacity dominates over free-free absorption, leading the latter 
to be neglected when calculating the disk's structure.

The equation of state concerning the sound speed $c_{\rm{s}}$ is 
\begin{equation}
\rho c_{\rm{s}}^2 = \rho V_{\rm{A}}^2 +
\frac{\rho k_{\rm{B}} T_{\rm{c}}}{\mu m_{\rm{p}}} +
\frac{1}{3}aT_{\rm{c}}^4\ . \label{eq11}
\end{equation}

Considering the viscosity prescription, we have
\begin{equation}
\frac{3}{2}\rho \nu \Omega_{\rm{K}} = \alpha P\ . \label{eq12}
\end{equation}

By solving the ten equations, Equations~(\ref{eq1}-\ref{eq8}), and (\ref{eq11}-\ref{eq12}),
 for the ten variables $V_{\rm{A}}$, $P$, $c_{\rm{g}}$, $c_{\rm{s}}$, $H$, 
$\Sigma$, $V_{\rm R}$, $\nu$, $\rho$ and $T_{\rm{c}}$ with given parameters 
$M_{\rm BH}$, $\alpha$, $\dot M$, $\lambda$, and $l$. We obtain the 
accretion timescale $t_{\rm{acc}} = -R/V_{\rm{R}}$. In the following 
calculations, we define as $\dot{M}_{\rm{Edd}} = 16L_{\rm{Edd}}/c^2 = 64\pi GM_{\rm{BH}}
/\kappa_{\rm{es}} c$, and fix $R_{\rm{out}} = 50R_{\rm{g}}$, where a typical region of 
the optical/UV emission.
 
\section{Numerical Results} \label{sec: numerical Results}

Before detailed numerical calculations, we do some analysis of
the basic characteristic timescales of the accretion disk, including 
the dynamical, thermal, and viscous timescales.
Their expressions are as follows 
\citep{1985apa..book.....F,2008bhad.book.....K},

\begin{equation}
t_{\rm{dyn}} \sim  \frac{1}{\Omega_{\rm{K}}}, \label{eq13}
\end{equation}

\begin{equation}
t_{\rm{th}} \sim \frac{1}{\alpha \Omega_{\rm{K}}}, 
\label{eq14}
\end{equation}

\begin{equation}
t_{\rm{vis}} \sim \frac{1}{\alpha \Omega_{\rm{K}}}\left(\frac{H_{\rm{SSD}}}{R}\right)^{-2}, \label{eq15}
\end{equation}

Regarding the observed timescales of CL-AGNs, $t_{\rm{dyn}}$ and 
$t_{\rm{th}}$ are too short, while $t_{\rm{vis}}$ is too long \citep[e.g.,][]
{2018ApJ...864...27S,2022arXiv221105132R}. 
From Equations ~(\ref{eq14}) and ~(\ref{eq15}), it is seen that 
$t_{\rm{th}} \propto (H_{\rm{SSD}}/R)^{0}$ and $t_{\rm{vis}} \propto (H_{\rm{SSD}}/R)^{-2}$. 
Some researchers  \citep[e.g.,][]{2018ApJ...864...27S,2022arXiv221105132R} have 
suggested that setting $t \propto (H_{\rm{SSD}}/R)^{-1}$ may help address these 
inconsistencies. When outflows are not considered (i.e., $l = 0,p = 0$), this 
corresponds to the case of magnetically dominated viscous disks. By using 
Equations ~(\ref{eq3}) and ~(\ref{eq6}), Equation ~(\ref{eq7}) can reduce to
\begin{equation}
V_{\rm{R}} = - \frac{3\nu f^{-1}g}{2R} \ , \label{eq16}
\end{equation}
then, the accretion timescale
\begin{equation}
t_{\rm{acc}} = -\frac{R}{V_{\rm{R}}}=\frac{2fg^{-1}}{3\alpha \Omega_{\rm{K}}} 
\left(\frac{H}{R}\right)^{-2} \propto \left(\frac{H}{R}\right)^{-2}\ . \label{eq17}
\end{equation}

For $\lambda = 1$, $V_{\rm{A}}\gg c_{\rm{g}}$, using 
Equations~(\ref{eq1}-\ref{eq2}) and~(\ref{eq5}),
we have $c_{\rm{s}}\approx V_{\rm{A}}$, then
\begin{equation}
t_{\rm{acc}} \propto \left(\frac{H}{R}\right)^{-2} \propto 
\left(\frac{c_{\rm{g}}}{V_{\rm{K}}}\right)^{-1}\propto 
\left(\frac{H_{\rm{SSD}}}{R}\right)^{-1}\ . \label{eq18}
\end{equation}
According to the above analysis, we have shown that $t_{\rm{acc}}
\propto (H/R)^{-2} \sim (H_{\rm{SSD}}/R)^{-1}$.
However, based on the following numerical calculations, we find that
$t_{\rm{acc}} \propto (H_{\rm{SSD}}/R)^{-1}$ cannot fully explain
the rapid variability of CL-AGNs. On the other hand, we propose that 
outflows play an important role in explaining the phenomenon of CL-AGNs.

\begin{figure}
\centering
\includegraphics[height = 12cm,width = 14cm]{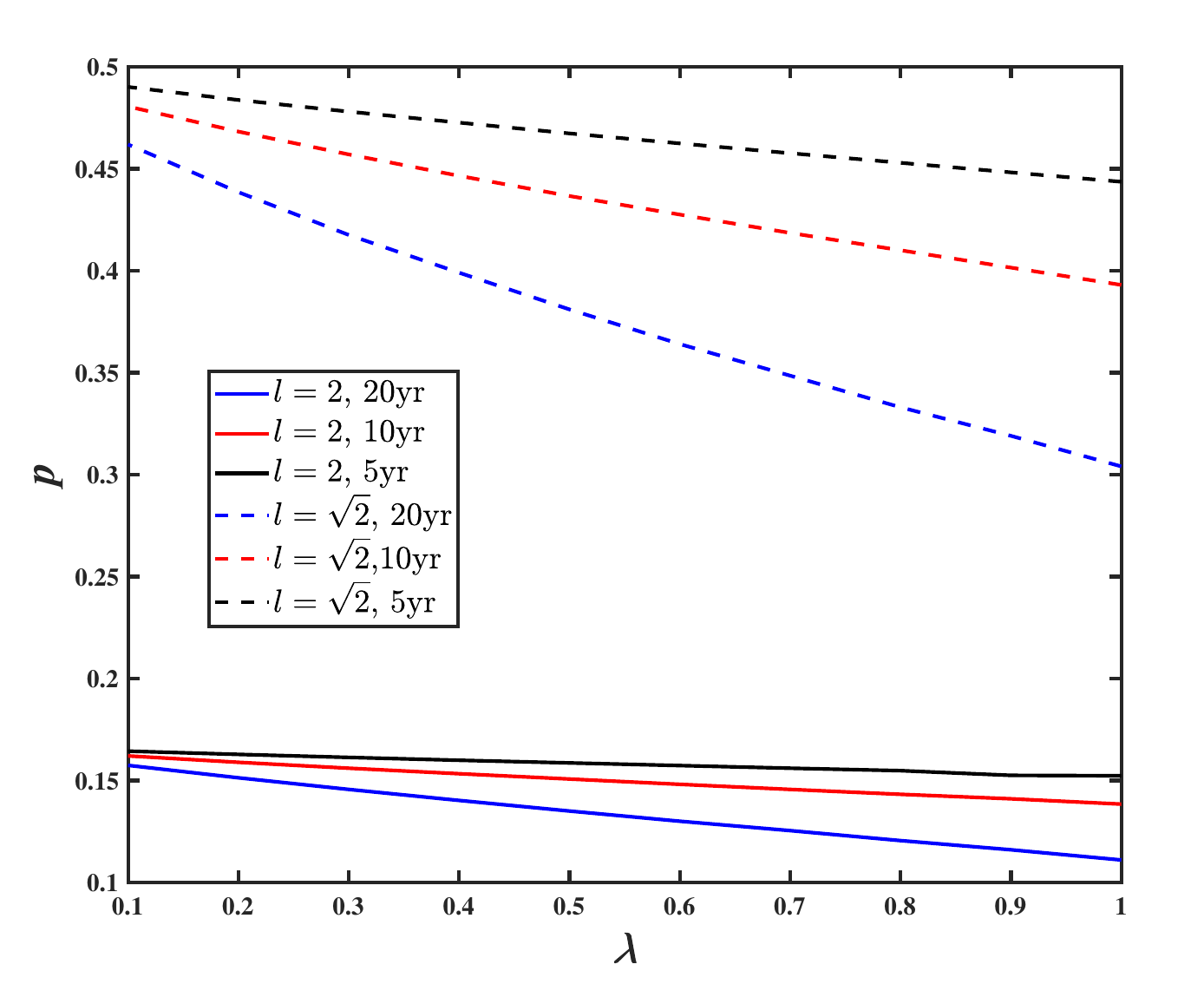}
\caption{Timescales are shown in the $\left\{p, \lambda \right\}$- plane for 
a given $L_{\rm{bol}}/L_{\rm{Edd}} = 0.05$. The dashed line represents 
$(l = \sqrt{2})$ and the solid line corresponds to $(l = 2)$. 
Each of the color lines corresponds to a specific value on the timescale. \label{fig:fig1}}
\end{figure}

Figure~\ref{fig:fig1} shows the relationship of outflow 
parameter $p$ and the strength of the turbulent (toroidal) magnetic 
field $\lambda$ for a given timescale when the $L_{\rm{bol}}/L_{\rm{Edd}} = 0.05$, 
where $M_{\rm{BH}} = 10^8 M_{\sun}$, $\alpha = 0.1$, and $L_{\rm{bol}}$ 
($=\int 2\pi R Q_{\rm{rad}}dR$) is the bolometric luminosity. The dashed line 
represents $(l = \sqrt{2})$ and the solid line corresponds to $(l = 2)$, 
We would like to stress that, the scaleheight of the disk
in our model (magnetically dominated viscous disks with outflows),
is significantly larger than that of a standard thin disk
(see Appendix ~\ref{sec: App} for the derivation of scaleheight).
Therefore, it is possible to further shorten the accretion timescale. The 
blue, red, and black lines represent $20,10$, and $5$ years, respectively. For 
comparison, we calculated the accretion timescale of the standard thin disks 
($\lambda = 0,\ l = 0,\ p = 0$, note that for standard thin disks: 
$\lambda \ll 1$. In this work, for simplicity, we followed the conventional 
approach $\lambda = 0$ for standard thin disks, as described in some
classic literature, e.g., Eq.(5.36) of \cite{2002apa..book.....F} and
Eq. (3.41) of \cite{2008bhad.book.....K}) 
and the magnetically dominated viscous disks 
($0 < \lambda \leq 1,\ l = 0,\ p = 0$) under the same conditions 
($\alpha = 0.1,\ L_{\rm{bol}}/L_{\rm{Edd}} = 0.05$): 
$t_{\rm{acc}} \sim 3.35 \times 10^3$ year and $t_{\rm{acc}} \geq 63.36$ 
year. Our calculations demonstrate that if the role of outflows is considered 
in the magnetic pressure-dominated disk, the accretion timescale can be 
shortened to several decades or a few years. 

\begin{figure}
\centering
\includegraphics[height = 12cm,width = 14cm]{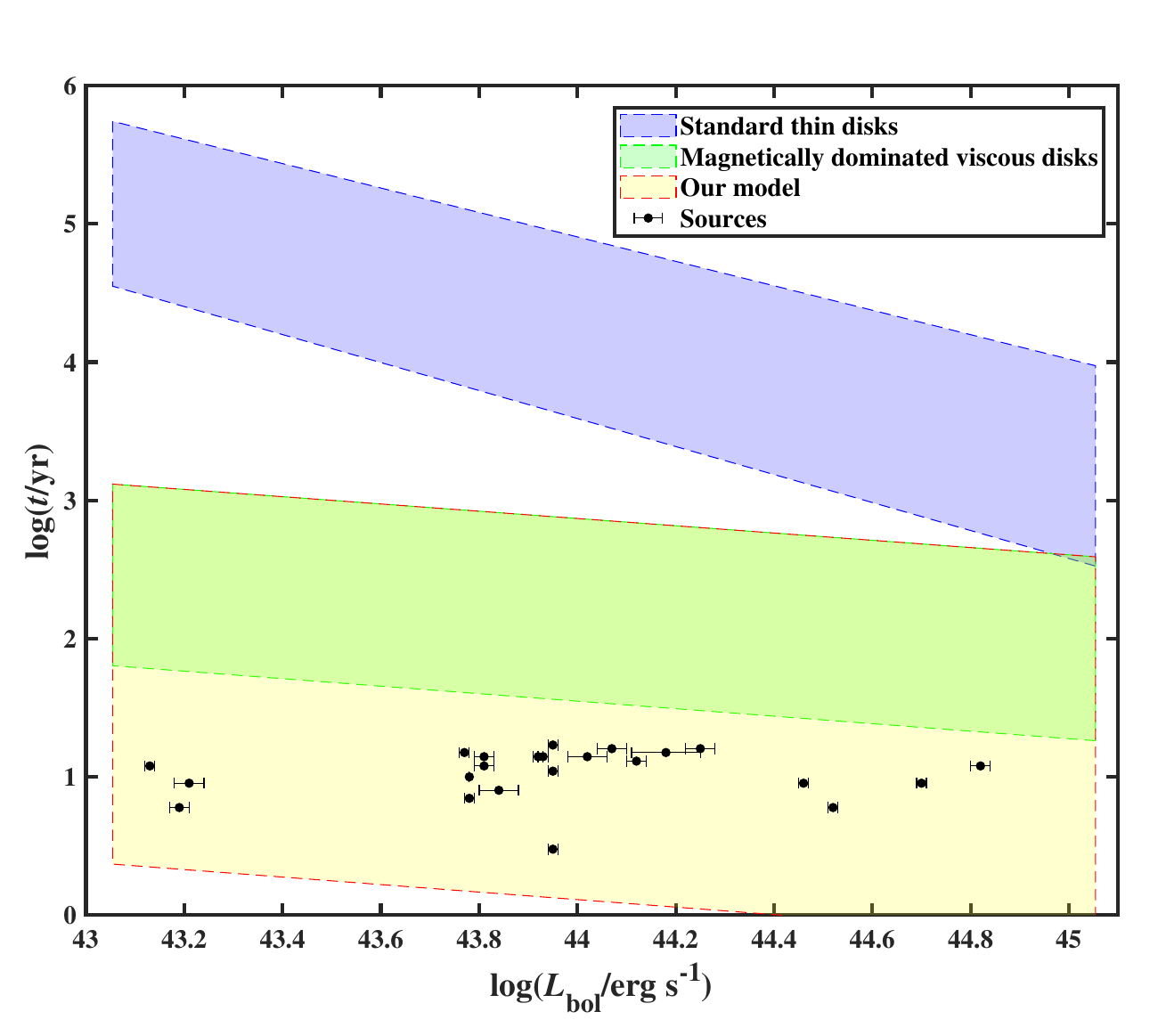}
\caption{The accretion timescale of accretion disks at $M_{\rm{BH}} = 10^8 M_{\sun}$ 
for different $\alpha$ and $L_{\rm{bol}}$. The regions of color correspond to 
the cases for the standard thin disks (blue), magnetically dominated 
viscous disks (green), and our model (magnetically dominated viscous 
disks with outflows, yellow). The black 
dots are the high state (turn-on) of CL-AGNs in Table~\ref{table}.  \label{fig:fig2}}
\end{figure}

To further understand the impact of the outflow, we adopt 
$M_{\rm{BH}} = 10^8 M_{\sun}$ and fix $\lambda = 1$ based on simulation results 
\citep{2023ApJ...945...57H}. We set $l = 2$, which corresponds to the case that 
significant angular momentum is carried away by outflows.
The range of the parameter $p$ is constrained by the given values
of $\lambda$ and $l$. In our case, the range $0 < p \leq 0.16$ corresponds
to $\lambda = 1$ and $l = 2$. In addition, by varying the viscosity parameter 
$\alpha \ (0.01 \sim 0.3)$
and the range of $L_{\rm{bol}}\ (0.001 \leq L_{\rm{bol}}/L_{\rm{Edd}} \leq 0.1)$,
we therefore obtain a decade range in accretion timescale in Figure~\ref{fig:fig2}.
The blue region is the accretion timescale of standard thin disks, the green 
region is the accretion timescale of magnetically dominated viscous 
disks, and the yellow region takes into account the role of outflows. 
Black dots denote the high state (turn-on) of CL-AGNs discovered by 
\cite{2022ApJ...926..184J} via SDSS, with 
physical quantities listed in Table~\ref{table}. Figure~\ref{fig:fig2} shows 
that the timescale of magnetically dominated viscous disks is 
significantly shorter than that of standard thin disks, but still longer than 
the observed timescale. However, when the angular momentum is carried 
away by the outflow, the radial velocity of the accretion disk can be 
significantly increased, as shown in Equation~(\ref{eq21}).
Thus, magnetically dominated viscous disks with outflows can further
shorten the accretion timescale and satisfy the rapid variability observed 
in CL-AGNs, as shown by the yellow region in Figure~\ref{fig:fig2}.

\begin{center} 
\begin{table*}[]
	\caption{Information about sources.}
	\begin{tabular}{lccccccccccccc}
		\hline
		\hline
		 \multicolumn{1}{c}{Name}  &  \multicolumn{2}{c}{$L_{\rm bol}$}   &   
$M_{\rm BH,vir}$  &  $\Delta t$ (yr) \\
		 &\multicolumn{2}{c}{$\rm log(erg \,s^{-1})$}  & ${\rm log}(M/M_{\odot})$ & {\rm yr} \\
		 & on & off & & \\
		 \multicolumn{1}{c}{(1)} & \multicolumn{2}{c}{(2)} &  (3)&(4)\\
		\hline
J0126-0839&	  $44.52\pm 0.01$ &	   $43.72\pm 0.03$ &	    $ 7.70^{+ 0.11}_{- 0.10} $ &	   6.0&\\
J0159+0033&	  $44.70\pm 0.01$ &	   $44.18\pm 0.01$ &	      $ 7.71^{+ 0.12}_{- 0.10} $ &	   9.0&\\
ZTF18aaabltn&	  $43.13\pm 0.01$ &	   $42.25\pm 0.11$ &	     $ 7.13^{+ 0.13}_{- 0.11} $ &	   12.0&\\
J0831+3646&	  $43.81\pm 0.02$ &	   $43.35\pm 0.05$ &	     $ 8.05^{+ 0.18}_{- 0.17} $ &	   14.0&\\
J0909+4747&	  $43.92\pm 0.01$ &	   $43.50\pm 0.01$ &	     $ 7.32^{+ 0.13}_{- 0.12} $ &	   14.0&\\
J0937+2602&	  $43.84\pm 0.04$ &	   $43.72\pm 0.01$ &	     $ 7.33^{+ 0.20}_{- 0.19} $ &	   8.0&\\
J1003+3525&	  $43.81\pm 0.03$ &	   $43.50\pm 0.02$ &	     $ 8.15^{+ 0.16}_{- 0.15} $ &	   12.0&\\
J1104+6343&	  $43.19\pm 0.02$ &	   $43.07\pm 0.06$ &	     $ 7.72^{+ 0.12}_{- 0.10} $ &	   6.0&\\
J1110-0003&	  $44.25\pm 0.03$ &	   $44.07\pm 0.01$ &	     $ 8.02^{+ 0.12}_{- 0.11} $ &	   16.0&\\
J1115+0544&	  $43.93\pm 0.01$ &	   $42.70\pm 0.07$ &	     $ 7.25^{+ 0.10}_{- 0.08} $ &	   14.0&\\
J1132+0357&	  $44.12\pm 0.02$ &	   $42.86\pm 0.04$ &	     $ 7.17^{+ 0.14}_{- 0.13} $ &	   13.0&\\
ZTF18aasuray&	  $43.95\pm 0.01$ &	   $42.54\pm 0.02$ &	     $ 7.24^{+ 0.16}_{- 0.15} $ &	   17.0&\\
ZTF18aasszwr&	  $45.20\pm 0.00$ &	   $43.01\pm 0.06$ &	     $ 8.68^{+ 0.10}_{- 0.09} $ &	   16.0&\\
ZTF18aahiqf&	  $43.77\pm 0.01$ &	   $42.70\pm 0.04$ &	     $ 8.26^{+ 0.15}_{- 0.13} $ &	   15.0&\\
J1259+5515&	  $44.02\pm 0.04$ &	   $43.48\pm 0.06$ &	     $ 7.39^{+ 0.11}_{- 0.10} $ &	   14.0&\\
J1319+6753&	  $44.07\pm 0.03$ &	   $43.81\pm 0.02$ &	     $ 7.17^{+ 0.18}_{- 0.17} $ &	   16.0&\\
J1358+4934&	  $43.95\pm 0.01$ &	   $43.35\pm 0.02$ &	     $ 7.15^{+ 0.13}_{- 0.12} $ &	   3.0&\\
J1447+2833&	  $44.46\pm 0.01$ &	   $44.03\pm 0.00$ &	     $ 7.46^{+ 0.12}_{- 0.10} $ &	   9.0&\\
ZTF18aajupnt&	  $42.04\pm 0.13$ &	   $41.92\pm 0.15$ &	     $ 6.45^{+ 0.11}_{- 0.10} $ &	   16.0&\\
J1533+0110&	  $43.78\pm 0.01$ &	   $43.49\pm 0.02$ &	     $ 7.38^{+ 0.11}_{- 0.10} $ &	   7.0&\\
J1545+2511&	  $43.95\pm 0.01$ &	   $43.70\pm 0.01$ &	     $ 7.34^{+ 0.11}_{- 0.10} $ &	   11.0&\\
J1550+4139&	  $44.18\pm 0.07$ &	   $43.53\pm 0.05$ &	     $ 7.25^{+ 0.10}_{- 0.08} $ &	   15.0&\\
J1552+2737&	  $43.21\pm 0.03$ &	   $42.84\pm 0.05$ &	     $ 7.94^{+ 0.12}_{- 0.11} $ &	   9.0&\\
J1554+3629&	  $44.82\pm 0.02$ &	   $42.93\pm 0.08$ &	     $ 8.00^{+ 0.15}_{- 0.14} $ &	   12.0&\\
PS1-13cbe&	    $43.78\pm 0.00$ &	   $42.54\pm 0.09$ &	        $ 6.89^{+ 0.10}_{- 0.08} $ &	   10.0&\\
		\hline
\end{tabular}
\tablecomments{Column 2: The bolometric luminosity of CL-AGNs ``turn-on" and 
``turn-off" (high and low states). Column 3: the virial black hole mass of 
CL-AGNs. Column 4: The lag between the ``turn-on" and ``turn-off" epochs in 
CL-AGNs. More physical quantities and derivation processes can be found in 
\cite{2022ApJ...926..184J}.}    \label{table}
\end{table*}
\end{center} 

\begin{figure}
\centering
\includegraphics[height = 12cm,width = 14cm]{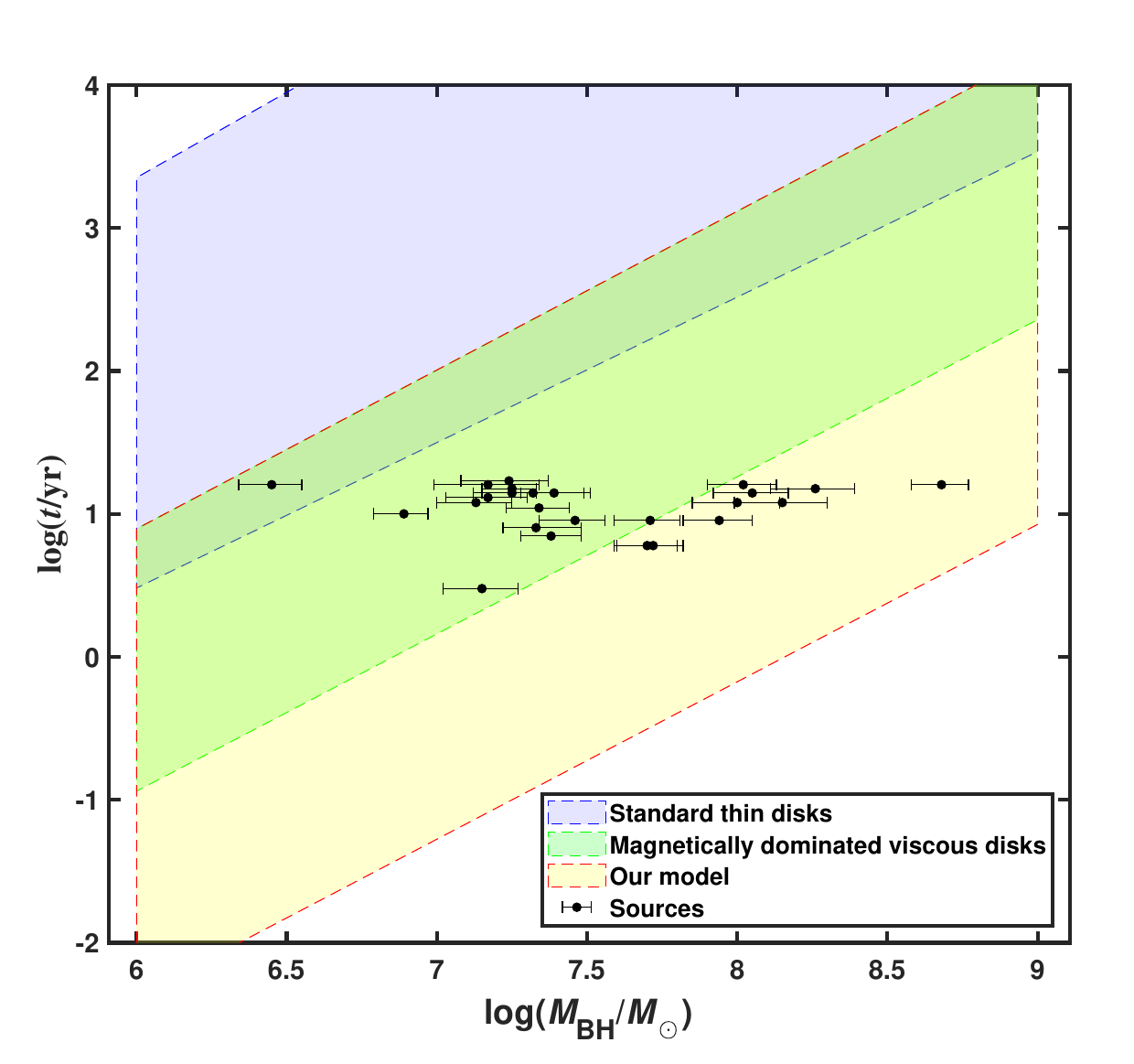}
\caption{The variations of the accretion timescale with mass under the 
$\lambda = 1, l = 2,\alpha = 0.01 \sim 0.3, 
L_{\rm{bol}}/L_{\rm{Edd}} = 0.001 \sim 0.1$. The meaning of the colors
and symbols is the same as in Figure~\ref{fig:fig2}.
For the standard thin disks, the accretion timescale is much larger 
than the observed timescale. Magnetically dominated viscous disks only 
explain part of the CL-AGNs. However, our model (magnetically dominated 
viscous disks with outflows) can naturally explain all the observed sources.  
\label{fig:fig3}}
\end{figure}

Based on the preceding analysis, our model predicts a timescale that can be as 
short as several years, which aligns with the observed timescale of CL-AGNs. 
However, this result is based on a fixed mass assumption and may not accurately 
reflect timescales for CL-AGNs with different masses. Figure~\ref{fig:fig3} 
illustrates the variations in accretion time with mass. 
Similar to Figure~\ref{fig:fig2}, we choose $\lambda = 1,\ l = 2,$
and $0 < p \leq 0.16$. By varying $\alpha \ (0.01 \sim 0.3)$
and $L_{\rm{bol}}\ (0.001 \leq L_{\rm{bol}}/L_{\rm{Edd}} \leq 0.1)$, as well as
a given range for $M_{\rm{BH}}\ (10^6 M_{\sun} \leq M_{\rm{BH}} \leq 10^9 M_{\sun})$,
we therefore obtain a decade range in accretion timescale. As depicted in Figure~\ref{fig:fig3}, 
while the magnetically dominated viscous disks model can satisfy 
CL-AGNs with masses less than $10^{7.5}M_{\sun}$, it is ineffective for higher 
masses. In contrast, our model can accommodate all sources' timescales. 
Therefore, the magnetic pressure-dominated disk model, with outflows taken into 
account, appears to be particularly promising for resolving the inconsistencies 
between observed timescales and disk theory.

\section{Conclusions and Discussion} \label{sec: Summary}

In this work, following the recent simulation results of
\citet{2023ApJ...945...57H}, we have investigated the accretion timescale 
of the magnetic pressure-dominated accretion disk.
Our results show that, even the timescale is significantly shorter than that
of the standard thin disk, it is still beyond the rapid variability of CL-AGNs.
Furthermore, by taking into account the role of outflows,
which were generally found in simulations and observations,
we have shown that the accretion timescale can be even shortened.
By the detailed comparison of the theoretical accretion timescale
with the observations,
we propose that the magnetic pressure-dominated disk with outflows
can be responsible for the rapid variability of CL-AGNs.

The effects of magnetic fields on accretion disks have been extensively 
studied, and some early research \citep{1995A&A...295..807F,1997A&A...319..340F,2019MNRAS.490.3112J} 
found that magnetic winds can significantly decrease the accretion timescale. 
This finding has been confirmed by numerical simulations \citep{2021A&A...647A.192J,2023arXiv230210226S}
and has been used to explain observed phenomena such as YSOs \citep{2008A&A...479..481C,2022A&A...667A..17M}
and X-ray binaries \citep{2006A&A...447..813F,2018A&A...615A..57M}. Full MHD 
models \citep{1997A&A...319..340F} relate various parameters, such as $p$ with 
disk magnetization, $\alpha$ with disk magnetization, and $l$ with $p$. 
However, these models do not fully consider the role of magnetic 
(turbulence) pressure. We would point out that our model includes
magnetic pressure and utilizes a simple parameterization, and it is worth
further investigation on the relationship among these parameters.

Quasi-periodic eruptions (QPEs) are a newly discovered type of changing-look 
phenomenon characterized by rapid and extreme X-ray bursts with mass 
$M_{\rm{BH}} \sim 10^{5-6} M_{\sun}$, and periods as short as a few hours 
\citep[e.g.,][]{2019Natur.573..381M,2020A&A...636L...2G,2021ApJ...921L..40C,2021Natur.592..704A}. 
However, these short timescales cannot be explained even at $R = 50R_{\rm{g}}$ 
and $\alpha = 0.3$ using our present one-zone model.
In order to interpret QPEs, a thicker disk may be necessary.
\cite{2015ApJ...809..118B} proposed a vertically 
stratified model with a larger characteristic scale height $H \sim R$ which 
agrees with \cite{2013ApJ...767...30B}'s vertically stratified, local shearing 
box ideal MHD. Subsequent numerical simulations \citep{2016MNRAS.460.3488S} 
have also confirmed this model. In addition, \cite{2013ApJ...767...30B} also 
shows that a magnetocentrifugal mechanism could excite a strong outflow. 
Therefore, our future work will focus on a new outflow model based on
these findings to explain the QPE phenomenon.

\begin{acknowledgments}
We thank Zhi-Xiang Zhang for helpful discussions, and the anonymous referee
for constructive suggestions that improved the paper.
This work was supported by the National Natural Science Foundation of China
under grants 11925301, 12033006, and 12221003.
\end{acknowledgments}

\appendix
\renewcommand{\appendixname}{Appendix ~\Appendix{section}}
\section{expression of scaleheight}\label{sec: App}

In this Appendix, we will derive the detailed expression of scaleheight, 
which is mentioned in Section $\ref{sec: numerical Results}$.

For $\lambda = 1, V_{\rm{A}}\gg c_{\rm{g}}$, according to 
Equations~(\ref{eq1}-\ref{eq2}) and~(\ref{eq5}), the scaleheight takes the form:
\begin{equation}
\frac{H}{R} = \left(\frac{c_{\rm{g}}}{V_{\rm{K}}}\right)^{1/2}\ . \label{eq20}
\end{equation}

Combining Equations~(\ref{eq3}) and (\ref{eq6}-\ref{eq7}), we have
\begin{equation}
V_{\rm{R}}  = -\frac{3}{2}\frac{\alpha f^{-1}g}{R}\left(1-\frac{l^2p}
{p+1/2}\right)^{-1}c_{\rm{g}}\ . \label{eq21}
\end{equation}

The surface density is
\begin{equation}
\Sigma = \frac{\dot{M} fg^{-1}}{3\pi \alpha c_{\rm{g}} R}\left(1-\frac{l^2p}
{p+1/2}\right)\ . \label{eq22}
\end{equation}

Defining $\dot{m}\equiv \dot{M}/
\dot{M}_{\rm{Edd}}$ and $x \equiv R/R_{\rm{g}}$, we obtain an expression for 
the optical depth through the disk:
\begin{equation}
\tau =\frac{16}{3} \frac{\dot{m}fg^{-1}c}{\alpha c_{\rm{g}}}
x^{-1}\left(1-\frac{l^2p}{p+1/2}\right)\ . \label{eq23}
\end{equation}

$ Q_{\rm{vis}}$ can be reduced to

\begin{equation}
Q_{\rm{vis}} = \frac{6\dot{m} c^5 fg}{\kappa_{\rm{es}} GM_{\rm{BH}}}x^{-1}(x-1)^{-2}
\left(1-\frac{l^2p}{p+1/2}\right)\ , \label{eq24}
\end{equation}

Due to $Q_{\rm{rad}} = 4P_{\rm{r}}c/\tau = Q_{\rm{vis}}$, 
we can get
\begin{equation}
P_{\rm{r}}  = \frac{8\dot{m}^2 c^5f^2}{\alpha \kappa_{\rm{es}} GM_{\rm{BH}}}x^{-2}(x-1)^{-2}\left(1-\frac{l^2p}{p+1/2}\right)^2c_{\rm{g}}^{-1}\ . \label{eq26}
\end{equation}

According to the radiation pressure, we have

\begin{equation}
T_{\rm c} = \left(\frac{3P_{\rm{r}}}{a}\right)^{1/4}\ . \label{eq27}
\end{equation}

For $P_{\rm{g}}\gg P_{\rm{r}}$, we have

\begin{equation}
c_{\rm{g}} = \left(\frac{k_{\rm{B}}T_{\rm c}}{\mu m_{\rm{p}}}\right)^{1/2}\ . \label{eq28}
\end{equation}

Using Equations (\ref{eq26}-\ref{eq28}), we get the expression 
of $c_{\rm{g}}$:
\begin{equation}
c_{\rm{g}} = \left(\frac{k_{\rm{B}}}{\mu}m_{\rm{p}}\right)^{4/9}\left[
\frac{24\dot{m}^2 c^5f^2}{a\alpha \kappa_{\rm{es}} GM_{\rm{BH}}}x^{-2}(x-1)^{-2}\left(1-\frac{l^2p}{p+1/2}\right)^2\right]^{1/9}\ . \label{eq29}
\end{equation}

Substituting Equation~(\ref{eq29}) into~(\ref{eq20}) to get the $H/R$:
\begin{equation}
\frac{H}{R} = 3.5\times 10^{-2} \left(\alpha m\right)^{-1/18}\dot{m}^{1/9}
x^{-13/36}\left(x-1\right)^{7/18}\left(1-\frac{l^2p}{p+1/2}\right)^{1/9}f^{1/9}\ , \label{eq30}
\end{equation}
where $m = M_{\rm{BH}}/(10^{8}M_{\rm{\sun}})$. This is similar to the case of 
\cite{2023arXiv230210226S} with $H/R = 0.03$.

For non-outflow cases ($l = 0, p = 0$, see Equation (19) of \cite{2007MNRAS.375.1070B}), we have
\begin{equation}
\frac{H}{R} = 3.5\times 10^{-2} \left(\alpha m\right)^{-1/18}\dot{m}^{1/9}
x^{-13/36}\left(x-1\right)^{7/18}f^{1/9}\ , \label{eq31}
\end{equation}

For $\lambda = 0,\ l = 0,\ p = 0$, our model returns to the standard thin model. 
Similarly, we can obtain (see, e.g., Equation (3.66) of \citep{2008bhad.book.....K})
\begin{equation}
\frac{H_{\rm{SSD}}}{R} = 2.4\times 10^{-3} \left(\alpha m\right)^{-1/10}
\dot{m}^{1/5}x^{-13/20}\left(x-1\right)^{7/10}f^{1/5}\ . \label{eq32}
\end{equation} 
By comparing Equation~(\ref{eq31}) with Equation~(\ref{eq32}),
we find $H_{\rm{SSD}}/R \sim (H/R)^{1.8}$, which is quite close to the simple
analytic result $H_{\rm{SSD}}/R \sim (H/R)^{2}$ (Equation~(\ref{eq18})).

\end{document}